\documentclass[pre,aps,superscriptaddress,fleqn,twocolumn]{revtex4}
\usepackage{amsmath,amssymb,graphicx}
\usepackage{times,mathptm} 
\makeatletter
\renewcommand{\@biblabel}[1]{\hspace*{2ex}}%
\makeatother

\begin{document}

\addtolength{\textheight}{3\baselineskip} 
\title{Prediction of site-specific amino acid distributions and limits of
       divergent evolutionary changes in protein sequences}

\author{Markus~Porto}
\affiliation{Institut~f\"ur~Festk\"orperphysik,
             Technische~Universit\"at~Darmstadt, Hochschulstr.~8,
             64289~Darmstadt, Germany}

\author{H.~Eduardo~Roman}
\altaffiliation[Present address:]{Dipartimento~di~Fisica,
                Universit\`a~di~Milano-Bicocca,
                Piazza~della~Scienza~3, 20126~Milano, Italy}
\affiliation{Dipartimento~di~Fisica and INFN, Universit\`a~di~Milano,
             Via~Celoria~16, 20133~Milano, Italy}

\author{Michele~Vendruscolo}
\affiliation{Department~of~Chemistry, University~of~Cambridge,
             Lensfield~Road, Cambridge~CB2~1EW, UK}

\author{Ugo~Bastolla}
\affiliation{Centro~de~Astrobiolog{\'\i}a~(INTA-CSIC),
             28850~Torrej\'on~de~Ardoz, Spain}

\date{March 26, 2004, revised October 15, 2004}

\begin{abstract}
We derive an analytic expression for site-specific stationary distributions of
amino acids from the Structurally Constrained Neutral (SCN) model of protein
evolution with conservation of folding stability. The stationary distributions
that we obtain have a Boltzmann-like shape, and their effective temperature
parameter, measuring the limit of divergent evolutionary changes at a given
site, can be predicted from a site-specific topological property, the principal
eigenvector of the contact matrix of the native conformation of the protein.
These analytic results, obtained without free parameters, are compared with
simulations of the SCN model and with the site-specific amino acid
distributions obtained from the Protein Data Bank. These results also provide
new insights into how the topology of a protein fold influences its
designability, i.e.\ the number of sequences compatible with that fold. The
dependence of the effective temperature on the principal eigenvector decreases
for longer proteins, a possible consequence of the fact that selection for
thermodynamic stability becomes weaker in this case.
\end{abstract}

\maketitle

\section{Introduction}

The reconstruction of phylogenetic distances from sequence alignments requires
the use of a model of protein evolution. Evolutionary models are also needed
for reconstructing phylogenetic trees in the framework of maximum likelihood
methods (Felsenstein, 1981). In this context, both the mutational process and
the selection on protein folding and function must be taken into account. It is
well known that the local environment of a protein site within the native
structure influences the probability of acceptance of a mutation at that site
(Overington et al., 1990). Nevertheless, structural biology still has a very
limited impact in studies of phylogenetic reconstruction, and the models
commonly used in these studies rely on substitution matrices that do not
consider the structural specificity of different sites. The most used
substitution matrices, such as JTT (Jones et al., 1992), are obtained
extrapolating substitution patterns observed for closely related sequences, and
they have low performances when distant homologs are concerned (Henikoff and
Henikoff, 1993).

To take into account the protein-level selection, it is necessary to consider
site-specific amino acid distributions within a protein family (Halpern and
Bruno, 1998). Similarly, the use of site-specific substitution matrices
improves substantially maximum likelihood methods for reconstructing
phylogenetic trees (Li\`o and Goldman, 1998; Koshi and Goldstein, 1998; Koshi
et al., 1999; Thorne, 2000; Fornasari et al., 2002).

In the studies mentioned above, site-specific constraints are obtained either
through simulations of a model of protein evolution or fitting parameters in a
maximum likelihood framework. In this work we derive an analytic expression,
without adjustable parameters, for site-specific amino acid distributions and
show that they are in very good agreement with simulations of a model of
protein evolution and with site-specific amino acid distributions obtained from
the Protein Data Bank.

The site-specific evolutionary patterns that we predict are associated with a
key topological indicator of native state topology, namely the principal
eigenvector of the contact matrix (Bastolla et al., 2004; Porto et al., 2004).
Our approach is based on the Structurally Constrained Neutral model (SCN model;
Bastolla et al., 1999, 2000a, 2002, 2003a, 2003b), where possible mutations are
tested for conservation of structural stability using a computational model of
protein folding (Bastolla et al., 1998, 2000b, 2001). This approach is similar
in spirit to the Structurally Constrained Protein Evolution model (SCPE; Parisi
and Echave, 2001), that uses a different criterion to assess the thermodynamic
stability. Notice that the SCN model does not assume independence at different
sites, and in fact we have recently shown that site-dependence is linked to a
rather debated feature of neutral evolution, the overdispersion of neutral
substitutions (Bastolla et al., 2003b).

Site-specific amino acid distributions with Boltzmann form, similar to those
that we derive here, have already been used in other studies of protein
evolution, for instance by Koshi and Goldstein (1998), Koshi et al.\ (1999) and
Dokholyan et al.\ (2001; 2002). As we will discuss, the main difference between
our approach and these previous models consists in the fact that our approach
allows to compute analytically the Boltzmann parameters, without the need of
any empirical sequence family.

As we have previously shown (Bastolla et al., 2003a), the site-specific
conservation patterns provided by the SCN model, and therefore also by the
present analytic formulation, match qualitatively structural conservation
patterns found in bioinformatics studies (Ptitsyn, 1998; Ptitsyn and Ting,
1999). Further, the site-specific amino acid distribution that we present here
also enables us to derive an estimate of the sequence entropy compatible with a
given fold, which is termed the `designability' of the fold (Li et al., 1998;
Helling et al., 2001). These results are in agreement with recent studies
suggesting that the designability can be inferred from the protein topology
alone (Koehl and Levitt, 2002; England and Shakhnovich, 2003).

\section{Background}

In the SCN model (Bastolla et al., 2002, 2003a), starting from the protein
sequence in the Protein Data Bank (PDB), amino acid mutations are randomly
performed and accepted according to a stability criterion based on an effective
model of protein folding (Bastolla et al., 2000b, 2001). We use a
coarse-grained representation of protein structures as a binary contact matrix
$C_{ij}$ with elements equal to one if $i$ and $j$ are in contact (at least one
pair of heavy atoms, one belonging to each amino acid, are less than $4.5 \,
\mbox{\AA}$ apart), and zero otherwise. The effective free energy for a
sequence ${\mathbf{A}}$ in configuration ${\mathbf{C}}$ is approximated by an
effective contact free energy function $E({\mathbf{A}},{\mathbf{C}})$,
\begin{equation}\label{eq:energy}
\frac{E({\mathbf{A}},{\mathbf{C}})}{k_{\mathrm{B}} T} =
 \sum_{i < j} C_{ij} \, U(A_i, A_j) \, ,
\end{equation}
where ${\mathbf{U}}$ is a $20\times 20$ symmetric matrix with $U(a,b)$
representing the effective interaction, in units of $k_{\mathrm{B}} T$, of
amino acids $a$ and $b$ when they are in contact; we use the interaction matrix
derived by Bastolla et al.\ (2001). The free energy function,
Eq.~(\ref{eq:energy}), assigns lowest free energy to the experimentally known
native structure against decoys generated by threading the sequence over
protein-like structures derived from the PDB. If gaps in the sequence-structure
alignment are allowed, one has to use a suitable gap penalty term.

For testing the stability of a protein conformation, we use two computational
parameters: (i)~The effective energy per residue,
$E({\mathbf{A}},{\mathbf{C}})/N$, where $N$ is the protein length, which
correlates strongly with the folding free energy per residue for a set of 18
small proteins that are folding with two-states thermodynamics (correlation
coefficient $0.91$; U.~Bastolla, unpublished result); (ii)~The normalized
energy gap $\alpha$, which characterizes fast folding model sequences (Bastolla
et al., 1998) with well correlated energy landscapes (Bringelson and Wolynes,
1987; Goldstein et al., 1992; Abkevich et al., 1994; Gutin et al., 1995; Klimov
and Thirumalai, 1996). A mutated sequence is considered thermodynamically
stable if both computational parameters are above predetermined thresholds
(Bastolla et al., 2003a).

The interaction matrix ${\mathbf{U}}$ can be written in spectral form as
$U(a,b) = \sum_{\alpha=1}^{20} \epsilon_{\alpha} \, u^{(\alpha)}(a) \,
u^{(\alpha)}(b)$, where $\epsilon_{\alpha}$ are the eigenvalues, ranked by
their absolute value, and ${\mathbf{u}}^{(\alpha)}$ are the corresponding
eigenvectors. The main contribution to the interaction energy comes from the
principal eigenvector ${\mathbf{u}}^{(1)}$, which has a negative eigenvalue
$\epsilon_1 < 0$, as $\epsilon_1 \, u^{(1)}(a) \, u^{(1)}(b)$ has correlation
coefficient $0.81$ with the elements $U(a,b)$ of the full matrix. It is well
known that hydrophobic interactions determine the most significant contribution
to pairwise interactions in proteins, so that the components of the main
eigenvector are strongly correlated with experimental hydropathy scales (Casari
et al., 1992; Li et al., 1997). Considering only this main component, we define
an approximate effective energy function $H({\mathbf{A}},{\mathbf{C}})$,
yielding a good approximation to the full contact energy,
Eq.~(\ref{eq:energy}),
\begin{equation}\label{eq:hydro}
\frac{H({\mathbf{A}},{\mathbf{C}})}{k_{\mathrm{B}} T} \equiv
\epsilon_1 \sum_{i < j} C_{ij} \, h(A_i) \, h(A_j)\, .
\end{equation}
We call ${\mathbf{h}}({\mathbf{A}}) \equiv {\mathbf{u}}^{(1)}({\mathbf{A}})$
the Hydrophobicity Profile (HP) of sequence ${\mathbf{A}}$ (Bastolla et al.,
2004). This is an $N$-dimensional vector whose $i$-th component is given by
$h(A_i) \equiv u^{(1)}(A_i)$. The 20 parameters $h(a)\equiv u^{(1)}(a)$ are
obtained from the principal eigenvector of the interaction matrix, and we call
them \textit{interactivity} parameters.

The HP provides a vectorial representation of protein sequences. In turn, a
convenient vectorial representation of protein structures may be obtained by
the principal eigenvector of the contact matrix ${\mathbf{C}}$, which we denote
by PE and we indicate by ${\mathbf{c}}$. This vector maximizes the quadratic
form $\sum_{ij} C_{ij} \, c_i \, c_j$ with the constraint $\sum_i c_i^2 = 1$.
In this sense, $c_i$ can be interpreted as the effective connectivity of
position $i$, since positions with large $c_i$ are in contact with as many as
possible positions $j$ with large $c_j$. All its components have the same sign,
which we choose by convention to be positive. Moreover, if the contact matrix
represents a single connected graph (as it does for single-domain globular
proteins), the information contained in the principal eigenvector is sufficient
to reconstruct the whole contact matrix (Porto et al., 2004).

For any given protein fold, identified through its PE, we can define the
optimal HP, ${\mathbf{h}}_{\mathrm{opt}}$, as the HP that minimizes the
approximate effective free energy, Eq.~(\ref{eq:hydro}), for fixed mean and
squared mean of the hydrophobicity vector, $\big< h \big> = N^{-1} \sum_i
h(A_i)$ and $\big<h^2 \big> = N^{-1} \sum_i h(A_i)^2$. We impose a condition on
the mean hydrophobicity, $\big< h \big>$, because, if a sequence is highly
hydrophobic, not only the native structure but also alternative compact
structures will have favorable hydrophobic energy. Selection to maintain a
large normalized energy gap is therefore expected to place constraints on the
value of $\big< h \big>$.

 From the above property of the PE, ${\mathbf{c}}$, it is clear that the
optimal HP, ${\mathbf{h}}_{\mathrm{opt}}$, is strongly correlated with the PE
(Bastolla et al., 2004). In this formulation, $\big< h \big>$ and $\big< h^2
\big>$ are not determined by the native structure but depend on the mutation
and selection process (see below).

The optimal HP constitutes an analytic solution to the sequence design problem
with energy function given by Eq.~(\ref{eq:hydro}), and an approximate solution
to sequence design with the full contact energy function,
Eq.~(\ref{eq:energy}). In the SCN evolutionary model, attempted mutations are
accepted whenever the effective free energy and the normalized energy gap
overcome predefined thresholds. Therefore, we do not expect that the optimal HP
is ever observed during evolution, but we do expect thermodynamically stable
sequences compatible with a given fold to have HPs distributed around the
optimal one. We have verified this prediction, finding that the HPs of
individual selected sequences are correlated with the principal eigenvector of
the protein fold with correlation coefficient of $0.45$ (averaged over seven
protein folds and over hundred thousands of simulated sequences per fold),
whereas the HP averaged over all sequences compatible with a given fold, $\big[
{\mathbf{h}} \big]_{\mathrm{evol}}$, correlates much more strongly with the
principal eigenvector of that fold, with mean correlation coefficient $0.96$
averaged over the same seven folds (Bastolla et al., 2004). These results show
that one can recover the optimal HP through an evolutionary average of the HPs
compatible with the protein fold.

We found a similar result for the protein families represented in the PFAM
(Bateman et al., 2000) and in the FSSP (Holm and Sander, 1996) databases. In
this case, however, the correlation between the principal eigenvector and the
evolutionary average of the HP is weaker: The average correlation coefficient
is $0.57$ for PFAM families and $0.58$ for FSSP families (Bastolla et al.,
2004). This weaker correlation is not unexpected, since functional
conservation, which is not accounted for in the SCN model, plays an important
role in protein evolution, and the effective energy function that we use to
test thermodynamic stability is only an approximation. In addition, the number
of sequences used to calculate the average HP in the PFAM and FSSP databases is
much smaller than the number of sequences obtained by the SCN model. When the
average HP is computed using only some hundreds of SCN sequences, the same
order of magnitude as in PFAM or FSSP families, the correlation with the
principal eigenvector also drops to values comparable to those observed for the
PFAM and the FSSP sequence databases.

\section{Results}

\subsection{Derivation of $\pi_i(a)$}

The results presented above suggest that the SCN model of protein evolution can
be represented as a trajectory in sequence space where the HP moves around the
optimal HP, which is strongly correlated with the principal eigenvector of the
protein fold's contact matrix (Bastolla et al., 2004). In this work, we use
this analogy to compute analytically the site-specific distribution of amino
acid occurrences $\pi_i(a)$, where $i$ indicates a protein position and $a$
indicates one of the $20$ amino acid types.

Our previous results indicate that the evolutionary average of the
hydrophobicity vector, $\big[ {\mathbf{h}} \big]_{\mathrm{evol}}$, is
correlated with the principal eigenvector ${\mathbf{c}}$ of the native contact
matrix. In order to derive an analytical expression for $\pi_i(a)$, we now
assume that the correlation coefficient between the principal eigenvector and
the average HP is $1$, i.e.\ $r\big( \big[ {\mathbf{h}} \big]_{\mathrm{evol}},
{\mathbf{c}} \big) = \big( \big< \big[ {\mathbf{h}} \big]_{\mathrm{evol}} \cdot
{\mathbf{c}} \big>- \big< \big[ h \big]_{\mathrm{evol}} \big> \big< c \big>
\big) / \big( \sigma_c \sigma_{[ h ]_{\mathrm{evol}}} \big) = 1$, and we obtain
\begin{equation}\label{eq:hevol}
\big[ h_i \big]_{\mathrm{evol}} \equiv \sum_{ \{ a \} } \pi_i(a) \, h(a) =
 A \left( c_i/\big< c \big> - 1 \right) + B \, ,
\end{equation}
where the sum over $\{ a \}$ is over all amino acids, and
\begin{equation}\label{eq:A+B}
A = \sqrt{\frac{\big< \big[ h \big]_{\mathrm{evol}}^2 \big> -
 \big< \big[ h \big]_{\mathrm{evol}} \big>^2}
 {\big( \big< c^2 \big> - \big< c \big>^2 \big)/\big< c \big>^2}}
\textrm{ and }
B = \big< \big[ h \big]_{\mathrm{evol}} \big> \, .
\end{equation}
We are representing here two kinds of average: The angular brackets denote the
average over the $N$ positions of the protein, $\big< f \big> = N^{-1} \sum_i
f_i$, and the corresponding standard deviation is denoted by $\sigma_f^2 =
\big< f^2 \big> - \big< f \big>^2$, whereas square brackets denote
position-specific evolutionary averages, $\big[ f \big]_{\mathrm{evol}}=\sum_{
\{ a \} } \pi_i(a) \, f(a)$.

Eqs.~(\ref{eq:hevol},\ref{eq:A+B}) are the conditions that the stationary
distributions $\pi_i(a)$ have to fulfill in order to guarantee a perfect
correlation between principal eigenvector and the average HP. We assume that
these conditions are the only requirement that the $\pi_i(a)$ have to meet, so
that at stationarity the $\pi_i(a)$ are the distributions of maximal entropy
compatible with the above conditions. It is well known that the solution of
this optimization problem are Boltzmann-like (exponential) distributions,
characterized by an effective `temperature' $|\beta_i|^{-1}$ that, in this
context, varies from site to site and measures the tolerance of site $i$ to
accept mutations over very long evolutionary times,
\begin{equation}\label{eq:distribution}
\pi_i(a) = \frac{\exp[-\beta_i \, h(a)]}%
 {\sum_{ \{ a' \} } \exp[-\beta_i \, h(a')]} \, ,
\end{equation}
with the constraint, Eq.~(\ref{eq:hevol}),
\begin{equation}\label{eq:constraint}
\sum_{ \{ a \} } \exp[-\beta_i \, h(a)]
 \left[ h(a) - A \left( c_i/\big< c \big> - 1 \right) - B \right] = 0 \, .
\end{equation}
Eq.~(\ref{eq:constraint}) states an analytical relation between the `Boltzmann
parameter' $\beta_i$ and the principal eigenvector component $c_i$, given the
two evolutionary parameters $A$ and $B$. One sees from this equation that
$\beta_i$ equals zero if $c_i/\big< c \big> = 1 + A^{-1} \big( \sum_{\{a\}}
h(a)/20 - \big< \big[ h \big]_{\mathrm{evol}} \big> \big)$, and $\beta_i$
becomes negative for larger $c_i$ and positive for smaller $c_i$. We expect
that the relationship between $\beta_i$ and $c_i$ is almost linear in the range
$\big| c_i/\big< c \big> - 1 \big| \ll 1$, whereas $\beta_i$ tends to minus
infinity when the average hydrophobicity at site $i$, $\big[ h_i
\big]_{\mathrm{evol}}$, tends to the maximally allowed value, and to plus
infinity when the average hydrophobicity at site $i$ tends to the minimum
allowed value.

Eq.~(\ref{eq:constraint}) can be interpreted as follows: (i)~Positions with
large eigenvector component $c_i$ are buried in the core of the protein and are
therefore with high probability occupied by hydrophobic amino-acids (positive
$h(a)$) and have large and negative $\beta_i$, (ii)~surface positions with
small $c_i$ are more likely occupied by polar amino acids (negative $h(a)$) and
have large and positive $\beta_i$, and (iii)~intermediate positions are the
most tolerant having a small negative or small positive $\beta_i$.

Similar forms of site-specific amino acid distributions have been previously
proposed by other authors, see for instance Koshi and Goldstein (1998), Koshi
et al.\ (1999) and Dokholyan et al.\ (2001; 2002). We will discuss analogies
and differences between these previous approaches and the present one in the
last section.

\subsection{Validation: SCN model}

We first verified these predictions using the simulated trajectories obtained
through the SCN model. We found that the site-specific stationary distributions
of amino acids, $\pi_i(a)$, are well fitted by an exponential function of
hydrophobicity, $\pi_i(a) \propto \exp[-\beta_i \, h(a)]$, where we use the
interactivity scale given by the main eigenvector of our interaction matrix.

Analytically, the expected site-dependent Boltzmann parameter $\beta_i$ can be
calculated in an implicit form by rewriting Eq.~(\ref{eq:constraint}) as
\begin{equation}\label{eq:analytical}
c_i/\big< c \big> = 1 + A^{-1} \left[
 \frac{\sum_{ \{ a \} } h(a) \exp[-\beta_i \, h(a)]}%
 {\sum_{ \{ a \} } \exp[-\beta_i \, h(a)]} - B \right] \, ,
\end{equation}
giving $c_i$ as a function of $\beta_i$. To obtain $\beta_i$ as a function of
$c_i$, Eq.~(\ref{eq:analytical}) has to be inverted numerically, and the
parameters $A$ and $B$ are obtained from the simulation data by
Eq.~(\ref{eq:A+B}).

\begin{figure}[t]
\centerline{\includegraphics[scale=0.53]{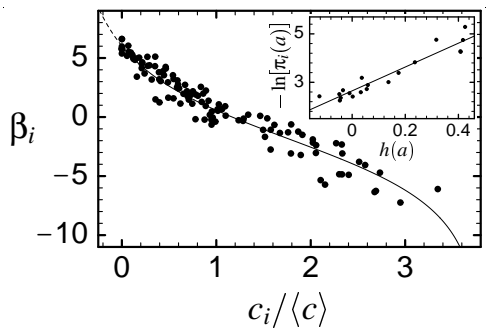}}
\caption{%
`Boltzmann parameter' $\beta_i$ as a function of the scaled principal
eigenvector component $c_i/\big< c \big>$ as obtained by the SCN model
for ribonuclease (PDB id.\ \texttt{7rsa}). The line shows the analytical
prediction, Eq.~(\ref{eq:analytical}), obtained using the mean
hydrophobicity $\big< \big[ h \big]_{\mathrm{evol}} \big> = 0.108$ and
the variance $\big< \big[ h \big]_{\mathrm{evol}}^2 \big> - \big<\big[ h
\big]_{\mathrm{evol}} \big>^2 = 0.0077$ as observed in the simulations of
the SCN model. The dashed part of the curve indicates the forbidden area
$c_i < 0$. The inset examplifies the numerical obtained $-\ln[\pi_i(a)]$
vs hydrophobicity $h(a)$ of amino acid $a$, as obtained for ribonuclease
at protein position $i=50$ (with $c_{50}/\big< c \big> = 0.124$),
yielding $\beta_{50} = 4.92$.
}
\label{fig:beta}
\end{figure}

\begin{table}[t]
\begin{tabular}{|l|l|c|c|c|c|}\hline
Protein & PDB id. & $N$ & $\beta$ & $R$ & $r$ \\
\hline
\hline
rubredoxin (mesophilic)		  & \texttt{1iro}	 & $53$  &
$0.98$ & $0.96$ & $0.96$ \\
rubredoxin (thermophilic) 	  & \texttt{1brf\_A} & $53$  &
$0.98$ & $0.96$ & $0.97$ \\
SH3 domain					  & \texttt{1aey}	 & $58$  &
$0.93$ & $0.97$ & $0.95$ \\
cytochrome~c  				  & \texttt{451c}	 & $82$  &
$0.94$ & $0.84$ & $0.84$ \\
ribonuclease  				  & \texttt{7rsa}	 & $124$ &
$0.94$ & $0.83$ & $0.89$ \\
lysozyme  					  & \texttt{3lzt}	 & $129$ &
$0.93$ & $0.73$ & $0.82$ \\
myoglobin 					  & \texttt{1a6g}	 & $151$ &
$0.95$ & $0.94$ & $0.88$ \\
ubiquitin conjugating enzyme    & \texttt{1u9a\_A} & $160$ &
$0.90$ & $0.88$ & $0.79$ \\
TIM barrel					  & \texttt{7tim\_A} & $247$ &
$0.94$ & $0.88$ & $0.84$ \\
kinesin						  & \texttt{1bg2}	 & $323$ &
$0.87$ & $0.79$ & $0.73$ \\
\hline
\end{tabular}
\caption{Correlation coefficients between site-specific quantities
predicted using the equations derived in this work and observed in SCN
simulations. Fourth column: `Boltzmann parameters' $\beta$. Fifth column:
Rigidities $R$. Sixth column: Substitution rates $r$. Note that sites
containing cysteine residues are not included in the calculation of the
correlation coefficients, as cysteine residues form pairwise disulphide
bridges (which are very poorly represented through hydrophobicity) and
are strictly conserved in our evolutionary model.}
\label{tab:summary}
\end{table}

The predictions derived in this way, for the ribonuclease fold (PDB id.\
\texttt{7rsa}), are compared in Fig.~\ref{fig:beta} to the $\beta_i$ obtained
by fitting the distributions $\pi_i(a)$ simulated through the SCN model. The
prediction does not involve any adjustable parameter, since $A$ and $B$ are
calculated from $\big< \big[ h \big]_{\mathrm{evol}} \big>$ and $\big< \big[ h
\big]_{\mathrm{evol}}^2 \big>$ as determined by the results of the SCN model.
The latter values do not depend on the native structure, but they depend
generally on the mutation and selection process, and specifically on its
realization as simulated through the SCN model. The other proteins studied show
a similar behavior, and the mean correlation coefficient between predicted and
observed $\beta_i$ is $0.935$ (cf.\ Table~\ref{tab:summary}).

\begin{figure}[t]
\centerline{\includegraphics[scale=0.53]{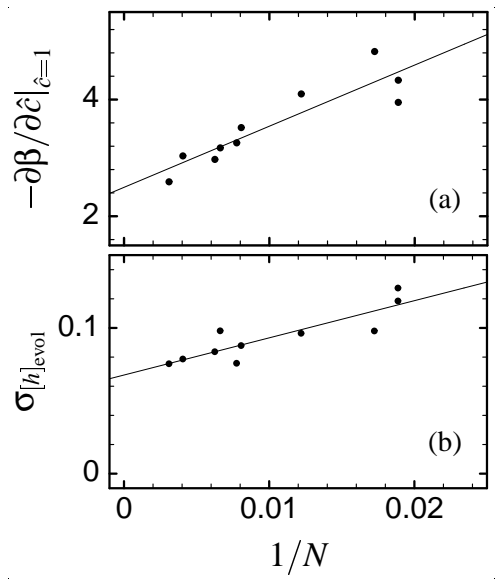}}
\caption{%
(a)~Plot of slope $-\partial \beta/\partial \hat{c}$, with $\hat{c}
\equiv c/\big< c \big>$ obtained at $\hat{c} = 1$, vs inverse chain
length $1/N$. The line shows a fit of the form $A_0 + A_1/N$. (b)~Plot of
the standard deviation of the hydrophobicity, $\sigma_{[ h
]_{\mathrm{evol}}}$, vs inverse chain length $1/N$ as obtained by the SCN
model. The line shows a fit of the form $B_0 + B_1/N$.
}
\label{fig:slope+sigmah}
\end{figure}

One notices from Fig.~\ref{fig:beta} that $\beta_i$ depends almost linearly on
the scaled principal eigenvector component $c_i/\big< c\big>$ over a wide range
of values. The slope of this linear part, the derivative $-\partial
\beta/\partial \hat{c}$ taken at $\hat{c} \equiv c/\big< c \big> = 1$, scales
with length $N$, and it can be well fitted by the expression $A_0 + A_1/N$ with
$A_0 = 2.50 \pm 0.21$ and $A_1= 104.8 \pm 9.2$ for the proteins listed in
Table~\ref{tab:summary}, see Fig.~\ref{fig:slope+sigmah}(a). One sees from this
expression that, as the protein length increases, the Boltzmann parameter
$\beta_i$ becomes less dependent on $c_i/\big< c \big>$ and therefore more
homogeneous. This result, at first sight unexpected, can be better understood
considering Eq.~(\ref{eq:analytical}), which implies that the derivative of
$\beta_i$ with respect to $\hat{c} \equiv c_i/\big< c \big>$ is proportional to
the standard deviation of the hydrophobicity,
\begin{equation}\label{eq:slope}
\frac{\partial\beta}{\partial \hat{c}}
\Big\vert_{\beta=0} \propto
 -\sqrt{\frac{\big< \big[ h \big]_{\mathrm{evol}}^2 \big> -
 \big< \big[ h \big]_{\mathrm{evol}} \big>^2}
 {\big( \big< c^2 \big> - \big< c \big>^2 \big)/\big< c \big>^2}} \, .
\end{equation}
The standard deviation of the hydrophobicity, measured from simulations of the
SCN model, decreases as a function of chain length as $B_0 + B_1/N$ with $B_0 =
0.068 \pm 0.016$ and $B_1 = 2.55 \pm 0.17$, see Fig.~\ref{fig:slope+sigmah}(b).
Thus, site-specificity decreases for longer proteins because the evolutionary
average of the HP becomes more homogeneous across different positions. This
result can be explained by noticing that not only the standard deviation, but
also the mean hydrophobicity, $\big< \big[ h \big]_{\mathrm{evol}} \big> $,
decreases as a function of chain length (notice again the two kinds of
average). So, the result that site-dependent Boltzmann parameters become more
homogeneous for longer proteins ultimately depends on the fact that longer
proteins tend to be less hydrophobic.

\subsection{Validation: PDB structures}

We compared our predictions to site-specific distributions obtained from a
representative subset of the Protein Data Bank (PDB). We considered a
non-redundant subset of single-domain globular proteins in the PDB, with a
sequence identity below 25\% (Hobohm and Sanders, 1994). Globularity was
verified by imposing that the fraction of contacts per residue was larger than
a length dependent threshold, $N_{\mathrm{c}}/N > 3.5 + 7.8 N^{-1/3}$. This
functional form represents the scaling of the number of contacts in globular
proteins as a function of chain length (the factor $N^{-1/3}$ comes from the
surface to volume ratio), and the two parameters were chosen so as to eliminate
outliers with respect to the general trend, which are mainly non-globular
structures. The condition of being single-domain was verified by imposing that
the normalized variance of the PE components was smaller than a threshold,
$\big( 1-N \big< c \big>^2 \big)/\big( N \big< c \big>^2 \big) < 1.5$.
Multi-domain proteins have PE components which are large inside their main
domains and small outside them. The PE components would be exactly zero outside
the main domains if the domains do not share contacts. Therefore, multi-domain
proteins are characterized by a larger normalized variance of PE components
with respect to single-domain ones. We have verified that the threshold of
$1.5$ is able to eliminate most of the known multi-domain proteins and very few
of the known single-domain proteins (data not shown).

We thus selected $774$ sequences of various lengths, $404$ of which were
shorter than $200$ amino acids. We first considered only this subset of short
proteins, and then the whole data set, divided in bins of similar lengths. We
counted the number of each of the 20 amino acids as a function of $c_i/\big< c
\big>$, where $\big< c \big>$ denotes the average over a single structure. We
used a bin-size of $0.05$ for $c_i/\big< c \big> \le 2.5$ and a bin-size of
$0.1$ for $c_i/\big< c \big> > 2.5$. Then, for each bin of $c_i/\big< c \big>$,
we fitted the observed distributions $\pi_{c_i/\left< c \right>}(a)$ with an an
exponential function of the hydrophobicity parameters, $\pi_{c_i/\left< c
\right>}(a) \propto \exp[-\beta_{c_i/\left< c \right>} \, h(a)]$. As in the
case of the SCN simulations, we used the interactivity scale derived from our
effective free energy function. The exponential fit was sufficiently good, and
yielded the observed Boltzmann parameters as a function of the normalized PE
components.

Next we calculated the predicted Boltzmann parameters through the equation:
\begin{equation}\label{eq:analytical2}
c_i/\big< c \big> = 1 + \tilde{A}^{-1} \left[
 \frac{\sum_{ \{ a \} } h(a) \exp[-\beta_{c_i/\left< c \right>} \, h(a)]}%
 {\sum_{ \{ a \} } \exp[-\beta_{c_i/\left< c \right>} \, h(a)]} -
\tilde{B} \right] \, ,
\end{equation}
with the parameters $\tilde{A}$ and $\tilde{B}$ now given by
\begin{equation}\label{eq:Aprime+Bprime}
\tilde{A} = \sqrt{
 \frac{\big< \big[ h \big]_{\mathrm{PDB}}^2 \big>_{c_i/\left< c \right>} -
 \big< \big[ h \big]_{\mathrm{PDB}} \big>_{c_i/\left< c \right>}^2}
 {\big[ \big( \big< c^2 \big> - \big< c \big>^2 \big)/
 \big< c \big>^2 \big]_{\mathrm{PDB}}}}
\textrm{ and }
\tilde{B} = \big< \big[ h \big]_{\mathrm{PDB}} \big>_{c_i/\left< c \right>} \, ,
\end{equation}
where the square brackets $\big[ h \big]_{\mathrm{PDB}}$ now denote, instead of
the evolutionary average over a protein family, the average over all positions
with fixed $c_i/\big< c \big>$, even belonging to different structures, whereas
angular brackets, $\big< \big[ h \big]_{\mathrm{PDB}} \big>_{c_i/\left< c
\right>}$, denote the average over all values of $c_i/\big< c \big>$. The
denominator $\big[ \big( \big< c^2 \big> - \big< c \big>^2 \big)/ \big< c
\big>^2 \big]_{\mathrm{PDB}}$ indicates the quantity $\big( \big< c^2 \big> -
\big< c \big>^2 \big)/ \big< c \big>^2$, obtained individually for each
structure, averaged over the whole set of structures.

\begin{figure}[t]
\centerline{\includegraphics[scale=0.53]{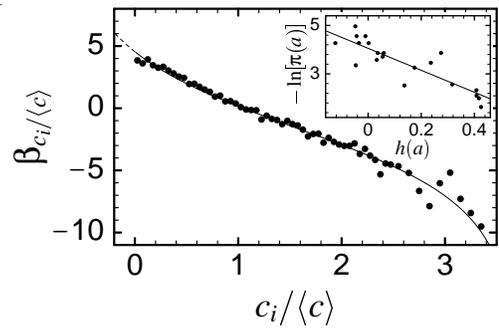}}
\caption{%
`Boltzmann parameter' $\beta_{c_i/\left< c \right>}$ as a function of the
scaled principal eigenvector component $c_i/\big< c \big>$ as obtained by
analyzing a subset of 404 non-redundant single-domain globular structures
of the PDB. The line shows the analytical prediction,
Eq.~(\ref{eq:analytical2}), obtained using the mean hydrophobicity $\big<
\big[ h \big]_{\mathrm{PDB}} \big> = 0.128$ and the variance $\big< \big[
h \big]_{\mathrm{PDB}}^2 \big> - \big<\big[ h \big]_{\mathrm{PDB}}
\big>^2 = 0.009$ as obtained from this set. The dashed part of the curve
indicates the forbidden area $c_i < 0$. The inset examplifies the
numerically obtained $-\ln[\pi(a)]$ vs hydrophobicity $h(a)$ of amino
acid $a$, as obtained for $c_i/\big< c \big> \in [ 2.45, 2.5 ]$, yielding
$\beta = -4.53$.
}
\label{fig:beta2}
\end{figure}

The observed Boltzmann parameters are compared in Fig.~\ref{fig:beta2} to the
predictions of Eq.~(\ref{eq:analytical2}). Notice that the agreement is indeed
remarkable, as the predictions do not involve any adjustable parameter, since
$\tilde{A}$ and $\tilde{B}$ are calculated from the PDB data.

\begin{figure}[t]
\centerline{\includegraphics[scale=0.53]{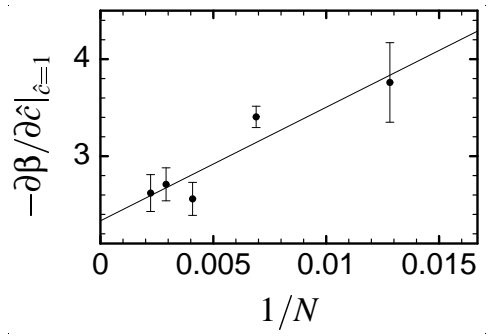}}
\caption{%
Plot of slope $-\partial \beta/\partial \hat{c}$, with $\hat{c} \equiv
c/\big< c \big>$ obtained at $\hat{c} = 1$, vs inverse chain length
$1/N$, obtained for structures of different length in the PDB.
}
\label{fig:beta_N}
\end{figure}

We stated in the previous section that, with SCN data, the dependence of the
Boltzmann parameters on the PE components becomes weaker, and the amino acid
distributions become more homogeneous, for longer chains. We tested whether
this interesting observation also holds for site-specific distributions
obtained from the PDB. To this end, we divided our set of proteins in five bins
of proteins with less than $100$, $200$, $300$, $400$, and $500$ amino acids,
respectively, and we measured the dependence of the Boltzmann parameters on the
PE component for each bin. From these data, we obtained the derivative of the
Boltzmann parameters at $\hat{c} \equiv c_i/\big< c \big> = 1$ simply by
fitting a straight line to $\beta(\hat{c})$ for $\hat{c}$ in the range between
$0.5$ and $1.5$. For all lengths considered, the fit was reasonably good, with
correlation coefficients always larger than $0.95$. The obtained slopes are
plotted against $1/N$ in Fig.~\ref{fig:beta_N}. Although the errors are rather
large, the data is compatible with the functional form $-\partial\beta/\partial
\hat{c}\approx A_0' + A_1'/N$. The parameters that we obtained by fitting this
relationship are also compatible, within error bars, with those obtained from
SCN data: We find, for PDB structures, $A_0' = 2.33 \pm 0.17$ and $A_1' = 117
\pm 25$. While a more careful study is still necessary on this issue, the
present results seem to confirm this length effect on the amino acid
distributions.

\subsection{Conservation and designability}

The results that we present imply that there is a direct relationship between a
structural indicator, the principal eigenvector of the contact matrix, and
site-specific measures of long-term evolutionary conservation and hence of
limits in divergent evolutionary changes. This relationship also provides a
link between the topology of a fold and its designability.

One convenient measure of the amino acid conservation at a given position is
given by the rigidity, defined as
\begin{equation}\label{eq:rigid}
R_i \equiv \sum_{ \{ a \} } \left[\pi_i(a) \right]^2
 = \frac{\sum_{ \{ a \} } \exp[-2 \beta_i \, h(a)]}%
 {\left\{ \sum_{ \{ a \} } \exp[-\beta_i \, h(a)] \right\}^2} \, .
\end{equation}
$R_i = 1$ means that the same amino acid is present at position $i$ in all
sequences, i.e.\ the conservation is total and $\beta_i^{-1}= 0$. In general,
the rigidity decreases with increasing temperature $|\beta_i|^{-1}$. One can
use Eqs.~(\ref{eq:analytical},\ref{eq:rigid}) to compute the rigidity from the
principal eigenvector. However, the dependence becomes clearer when fitting
directly the rigidity as a function of the principal eigenvector component. We
did this on the data generated through the SCN model, finding that the best fit
is parabolic and the three parameters depend on the protein length. Instead of
the PE component $c_i$, we use the rescaled variable $x_i \equiv \sqrt{N} c_i$,
which is normalized such that its mean squared value is one, i.e.\ $\big< x^2
\big> = N^{-1} \sum_i x_i^2 = 1$. Using this scaled variable, the rigidities of
all proteins in our data set can be fitted by the expression
\begin{equation}\label{eq:rigid_fit}
R_i^{\mathrm{pred}} = A +
 \frac{B}{N^d} \left(x_i - C \big< x \big> - \frac{D}{\sqrt{N}} \right)^2 \, ,
\end{equation}
with $A = 0.0630 \pm 0.0017$, $B = 0.28 \pm 0.16$, $C = 1.879 \pm 0.102$, $D =
-5.36 \pm 0.89$, and $d = 0.47 \pm 0.11$. The factor $N^{-d}\approx N^{-1/2}$
can be interpreted as a further indication that rigidities tend to become more
homogeneous for larger proteins, as previously noted concerning the Boltzmann
parameters. Eq.~(\ref{eq:rigid_fit}) predicts the rigidity for all the proteins
that we studied with the SCN model, with an average root mean square relative
error of around 10\% and an average correlation coefficient between predicted
and observed rigidities of $0.88$ (cf.\ Table~\ref{tab:summary}). These values
change very little in leave-one-out tests, where we discard one protein for
estimating the parameters and then use it as a blind test.

A standard information-theoretic measure of site-specific sequence conservation
is given by the entropy of the amino acid distribution,
\begin{equation}
S_i \equiv -\sum_{\{ a \}} \pi_i(a) \log\left[ \pi_i(a) \right]
 = \log\left[ Z(\beta_i) \right] + \beta_i \left[ h_i \right]_{\mathrm{evol}} \, ,
\end{equation}
where $Z(\beta_i) \equiv \sum_a \exp(-\beta_i \, h(a))$. The entropy attains
its maximum value, $S_i = \log(20)$, at $\beta_i = 0$ and it decreases with
increasing $|\beta_i|$. Predictions of the entropy based on a different
approach, Eq.~(\ref{eq:distribution2}), using aligned protein families have
been obtained by Dokholyan et al.\ (2001; 2002).

An important property of the entropy is that its exponential, $\exp(S_i)$,
provides an estimate of the average number of amino acid types acceptable at
position $i$ over very long evolutionary times. The exponential of the sum of
all site-specific entropies, $\exp(\sum_i S_i)$, gives an estimate of the
sequence space compatible with a given fold, termed the designability of the
fold, where we are assuming that the amino acid distributions at different
positions are independent. Although this assumption is clearly oversimplified,
the estimate of designability that can be obtained should be a valuable
approximation, and our approach allows to connect it explicitly to a
topological feature of the protein native structure (Koehl and Levitt, 2002;
England and Shakhnovich, 2003).

\section{Discussion and Conclusions}

We have predicted analytically that amino acids at specific positions are
distributed according to a Boltzmann law. In the latter, the role of energy is
played by the hydrophobicity, measured through an \textit{interactivity} scale,
and that of temperature by a quantity strongly correlated with a structural
indicator, namely the corresponding component of the principal eigenvector of
the native contact matrix (PE). This prediction, which does not involve any
free parameter, is in very good agreement both with simulations of the SCN
model of protein evolution and with site-specific amino acid distributions that
we obtained from a representative subset of single-domain globular proteins in
the PDB.

The relationship between a structural profile (the PE) and an evolutionary
profile (the Boltzmann parameters) that we find has an interesting
interpretation. Positions with a large principal eigenvector component are
strongly interacting with other positions in the core of the protein.
Therefore, they preferentially host strongly hydrophobic amino acids (large
negative $\beta_i$). On the other hand, positions with small PE are contained
in loops and, with higher probability, they host polar amino acids (large
positive $\beta_i$).

Despite this general interpretation, we have verified, using the SCN model of
protein evolution, that the PE has the strongest predictive power among several
similar structural indicators that quantify the difference between core and
surface positions of a protein. As alternative structural indicators, we
considered the total number of contacts, the number of long-range contacts
(separated by more than $10$ residues along the sequence) and the contact order
(the average loop length of each contact involving a given site), and we
correlated them with measures of site-specific conservation, but we always
found a correlation significantly weaker than for the PE.

The distributions derived here refer to very long evolutionary times, when
memory of the starting sequence has been lost. We recall the assumptions that
we made for deriving the site-specific distributions. The first assumption is
that selection on folding stability can be represented effectively as a maximal
correlation between the hydophobicity profile (HP) of sequences compatible with
a given fold and the optimal HP of that fold, the latter basically coinciding
with the PE. This assumption follows directly from an approximation of our
effective free energy function with its principal (hydrophobic) component. The
second assumption is that the average of the HP of selected sequences over very
long evolutionary times has a correlation coefficient of unity with the PE,
i.e.\ all other energetic contributions average out. The third assumption is
that this correlation is the only relevant property of the site-specific amino
acid distributions, in other words, these distributions are the distributions
of maximal entropy whose site-specific averages have correlation of one with
the PE, thus fulfilling the stability requirement. From these three assumptions
the Boltzmann form of the amino acid distributions is straightforward. In order
to compute the site-specific Boltzmann parameters, however, we still have to
know the positional mean and standard deviation of the site-specific HPs. These
quantites are determined by the mutation process and by the selection
parameters. They can be computed directly from the data, in such a way that the
analytic prediction does not contain any free parameter.

Boltzmann distributions, as those proposed in this work, have a long history in
studies of protein structure and evolution. Structural properties of native
protein structures, as for instance amino acid contacts, have been assumed to
be Boltzmann-distributed (Miyazawa and Jernigan, 1985), and Boltzmann
statistics for structural elements was predicted in stable folds of globular
proteins (Finkelstein et al., 1995). Our work may point out to an alternative
or complementary explanation for such distributions.

Shakhnovich and Gutin (1993) proposed a model of sequence design through Monte
Carlo optimization, which produced a Boltzmann distribution in sequence space.
A mean field approximation of this model (Dokholyan et al.\ 2001; 2002) results
in site-specific amino acid distributions of the form
\begin{equation}\label{eq:distribution2}
\pi_i(a) \propto \exp[-\beta \, \phi_i(a)],
\end{equation}
formally similar to Eq.~(\ref{eq:distribution}). There are, however, three
important differences between our formulation and Eq.~(\ref{eq:distribution2}).
First, Eq.~(\ref{eq:distribution2}) was derived as an independent site
approximation to a Boltzmann distribution for entire sequences, whereas we
derived Eq.~(\ref{eq:distribution}) from the relationship between average
hydrophobicity at a given site and the PE component. Second, in
Eq.~(\ref{eq:distribution2}), the Boltzmann parameter $\beta$ is the same for
all sites, whereas we obtain the $\beta_i$ profile along the protein structure
from the PE. Third and most important, in order to compute
Eq.~(\ref{eq:distribution2}), Dokholyan et al.\ (2001; 2002) used aligned
families of natural proteins, whereas our computation only needs the PE and two
empirical values, the average and the standard deviation along the
hydrophobicity profile.

Koshi and Goldstein (1998) and Koshi et al.\ (1999) assumed site-dependent
Boltzmann distributions of physico-chemical amino acid properties, deriving
from it a protein evolution model that they use for phylogenetic reconstruction
in a maximal likelihood framework. Since the properties they used are
hydrophobicity and amino acid size, their proposed distributions are formally a
general case of those derived in this work. There are, however, two important
differences between their approach and ours: First, in our approach sites are
structurally classified according to the PE component, which is a structural
indicator very strongly correlated with conservation; Second, whereas the
Boltzmann parameters are treated as free parameters by Koshi et al., and fitted
in a maximal likelihood framework, in our approach they are computed
analytically.

Kinjo and Nishikawa have very recently pointed out the existence of a strong
relationship between hydrophobicity and the main eigenvector of substitution
matrices derived from protein alignments with various values of the sequence
identity of the aligned proteins (Kinjo and Nishikawa, 2004). They looked at
the eigenvector corresponding to the largest eigenvalue (in absolute value) of
the substitution matrices. For high sequence identities (above 35 percent),
this eigenvector indicates the propensity of the amino acid to mutate over
short evolutionary times (mutability). For low sequence identities (below 35
percent), corresponding to long evolutionary times, this eigenvector is very
strongly correlated with hydrophobicity. This correlation is easily understood
in light of the result presented here. In fact, Kinjo and Nishikawa used the
Henikoff's method (Henikoff and Henikoff, 1992) for deriving substitution
matrices from observed frequencies of aligned amino acids at positions with
various PE values. Using our notations, these substitution matrices can be
indicated as $M(a,b) \approx \log\left[\big< \pi_i(a) \, \pi_i(b)\big > /\big <
\pi_i(a)\big > \big < \pi_i(b)\big > \right]$, where the angular brackets
denote positional average. In other words, these substitution matrices measure
the tendency of two residue types $a$ and $b$ to co-occur at the same sites.
The relationship between large time substitution matrices and hydrophobicity
gives therefore independent support to our main result.

In dealing with protein families generated from natural evolution, our approach
has two main difficulties: First, it can not take into account functional
conservation; second, a large number of sequence pairs separated by very long
evolutionary time is required. Because of these difficulties, the site-specific
conservation predicted by our model agrees only qualitatively with
site-specific conservation observed in aligned protein families (Bastolla et
al., 2003a). For tackling these difficulties, we tested our analytic
predictions on a large set of proteins in the Protein Data Bank. Positions
belonging to different protein folds were counted together in the same
structural class characterized by the value of the normalized PE.

The agreement between the analytic prediction and the observed distributions
may be surprising in view of the fact that Eq.~(\ref{eq:analytical2}) takes
into account neither the genetic code nor the codon usage. It is known that the
observed frequencies of amino acids correlate strongly with their number of
synonymous codons, and even more strongly with their cumulative codon
frequencies expected in the simple hypothesis that the three bases constituting
each codon are independently distributed (Sueoka, 1961). Moreover, the
frequency of nucleotides at coding positions correlate strongly with the
frequency of nucleotides at non-coding positions such as the synonymous 3rd
codon positions (Bernardi and Bernardi, 1986), which are thought to reflect the
mutational process. On the other hand, mutations are neither adequately
represented in the SCN model, which is a model at the amino acid level where
all amino acid mutations are equiprobable.

The most straightforward way to include the biases of the mutational process is
to modify Eq.~(\ref{eq:distribution}) with $\pi_i(a)\propto w(a) \exp[-\beta_i
\, h(a)]$, where the weights $w(a)$ are either the number of synonymous codons
or their cumulative frequency, estimated using average nucleotide compositions.
We will address this issue in future work.

We also observed a length effect in the dependence of the Boltzmann parameters
$\beta_i$ on the PE. For longer proteins this dependence becomes weaker,
approximately following a law of the type $A_0 + A_1/N$. This result holds for
SCN simulations and, although less significantly, for frequency distributions
obtained from the PDB. Formally, the result directly follows from the fact that
the evolutionary average of the hydrophobicity, $\big[ h_i
\big]_{\mathrm{evol}}$, varies less and less across different positions $i$ for
longer proteins. This length dependence of the HP is consistent with the fact
that the position average of $\big[ h_i \big]_{\mathrm{evol}}$ decreases with
chain length, i.e.\ the length of a protein influences its composition (White,
1992; Bastolla and Demetrius, submitted), such that longer proteins tend to
become less hydrophobic. (We note, however, that these results were obtained
using the interactivity scale, and they can not be significantly generalized to
other hydropathy scales.) These results can be understood considering that
longer proteins are stabilized by a larger number of interactions per residue.
Selection for protein stability is therefore expected to be weaker on
individual interactions. In support of this prediction, the average interaction
energy per contact in PDB structures was found to decrease with chain length
(Bastolla and Demetrius, submitted).

The site-specific amino acid frequencies that we derived may find applications
in the estimation of site-specific substitution matrices (Li\`o and Goldman,
1998; Thorne, 2000; Fornasari et al., 2002), and site-specific structural
conservation. Our approach may provide analytic understanding of the limits
imposed by structural and functional contraints to the divergent evolution of
protein sequences and of the reliability of phylogenetic reconstructions from
protein sequences for anciently diverged taxa (Meyer et al., 1986). A better
understanding of structural conservation may also improve the prediction of
functional conservation at sites that appear more conserved than expected on a
structural basis alone (Casari et al., 1995; Ota et al., 2003). The dependence
of amino acid distributions on the PE that we presented here is also a step
towards determining the structural determinants of protein `designability' (Li
et al., 1998; Helling et al., 2001; England and Shakhnovich, 2003), namely the
volume of sequence space compatible with a given fold.

\acknowledgments

MP gratefully acknowledges financial support by the guest program of the
Max-Planck-Institut f\"ur Physik komplexer Systeme in Dresden, Germany, during
the early stages of this project. UB is sponsored by the I3P program of the
Spanish CSIC co-financed by the European Social Fund. We gratefully acknowledge
discussions with Alfonso Valencia, Julian Echave, and Gustavo Parisi, and
correspondence with Akira Kinjo.

\flushleft

\end{document}